\documentclass[12pt]{article}
\usepackage[T2A]{fontenc}
\usepackage[utf8]{inputenc}
\usepackage{setspace,amsmath}
\usepackage{bbold}
\usepackage{amsfonts}
\usepackage{amssymb}
\usepackage{amsthm}
\usepackage{cite}
\usepackage{fontenc}
\usepackage{feynmp-auto}
\usepackage[dvips]{graphicx}
\usepackage{cite}
\usepackage{float}
\usepackage[unicode,linktocpage=true,plainpages=false,pdfpagelabels=false]{hyperref}
\newtheorem{theorem}{Theorem}
\newtheorem{corollary}{Corollary}
\newtheorem{lemma}{Proposition}
\newtheorem{definition}{Definition}

\begin{document}
\title{Diagram technique for the heat kernel of the covariant \\ Laplace operator}
\author{A.~V.~Ivanov\thanks{E-mail: regul1@mail.ru}\\
{\it St. Petersburg Department of Steklov Mathematical 
Institute of}\\ {\it Russian Academy of Sciences, 27 Fontanka, 
St. Petersburg, Russia}}
\date{\vskip 15mm}
\maketitle
\begin{abstract}
We present a diagram technique used to calculate the Seeley–DeWitt coefficients for a covariant Laplace operator. We use the combinatorial properties of the coefficients to construct a matrix formalism and derive a formula for an arbitrary coefficient.
\end{abstract}

\newpage
\section{Introduction}

\textbf{Research motivations.} The Fock proper-time method described in \cite{1} was first used to work with Green’s functions but then found application in the field of gauge invariance \cite{11} and quantum gravitation \cite{12,13,14}. This approach is currently a rather convenient tool in mathematical physics \cite{22} and especially theoretical physics, where it is used in loop computations and in finding divergences and renormalization. As an example, we mention the computations of two and three loops for the Yang–Mills theory in the background field method \cite{2,3}. 
Computing the coefficients of the heat kernel expansion in the power series in the proper time (Seeley–DeWitt coefficients) is a separate labor-consuming problem. In the case of a second- order differential operator, answers have been obtained up to the power $\tau^{5-\frac{d}{2}}$, where $d$ is the space dimension \cite{15,16,17,18,19,20}. The most general survey of this field can be found in \cite{4}.\\

\noindent Our main subject here is the covariant Laplace operator in a space with a flat metric. In the framework of the general theory, we consider its heat kernel, whose coefficients in the power series expansion in the proper time satisfy relations (\ref{111}).
These differential equations admit solutions
(\ref{b19}) and (\ref{b36}), but such a form of the solutions is inconvenient when calculating the trace is required. Our main goal here is to study the combinatorial properties of the Seeley–DeWitt coefficients based on the developed diagram technique. This paper generalizes and mathematically justifies the formalism proposed in \cite{8}.\\

\noindent The covariant perturbation theory \cite{23,24,25}, where the expansion of the operator exponential is also used, is closest to our proposed approach. But the authors of \cite{23,24,25} were not interested in the combinatorics in higher orders of the perturbation theory. The higher orders of the heat kernel expansion were obtained in \cite{26} when determining the restrictions on the form of the gauge field strength. We do not impose such restrictions here.\\

\noindent\textbf{Results of the paper.}
This paper consists of two basic parts. The main result in the first part is the development of a diagram technique for working with the heat kernel of the covariant Laplace operator. In Definitions 2--4, we introduce the basic notions of this diagram technique. We also prove a theorem on differentiating a diagram, which plays the key role in computations. Further, as an example, we obtain the first coefficients arising in calculations with the determinant in the Yang–Mills theory \cite{5,6}.
In the second part of the paper, we derive the formula for an arbitrary Seeley–DeWitt coefficient $a_n(x,x)$ by using the matrices and operators acting on matrices. Our main result is contained in formulas 
(\ref{rt9}), (\ref{rt11}) and (\ref{ra7}). In conclusion, we calculate the coefficient $a_3(x,x)$ of the power $\tau^{3-d/2}$.

\section{Heat kernel}
The fundamental solution for the operator $A_0=-\partial^{\mu}\partial_{\mu}$ satisfies the equation
\begin{equation}
\label{a1}
A_0G_0(x,y)=\delta(x-y).
\end{equation} 
To obtain $G_0(x,y)$ by the heat kernel method \cite{1}, we must determine the function $K_0(x,y;\tau)$ that solves
the problem
\begin{equation}
\label{a2}
\left(\frac{\partial}{\partial\tau}+A_0\right)K_0(x,y;\tau)=0,\,\,\,\,\,\,
K_0(x,y;0)=\delta(x-y).
\end{equation}
Then
\begin{equation}
\label{a3}
G_0(x,y)=\int_{0}^{\infty}d\tau\,K_0(x,y;\tau).
\end{equation}
If we consider the Laplace operator $A_1=A_0+V(x)$ with a sufficiently good potential, then solving the
problem 
\begin{equation}
\label{a4}
\left(\frac{\partial}{\partial\tau}+A_1\right)K_1(x,y;\tau)=0,\,\,\,\,\,\,
K_1(x,y;0)=\delta(x-y),
\end{equation}
we obtain the corresponding fundamental solution
\begin{equation}
\label{a5}
G_1(x,y)=\int_{0}^{\infty}d\tau\,K_1(x,y;\tau).
\end{equation}
By the general theory, for a certain class of potentials, we can write the formula for the logarithm of the determinant
\begin{equation}
\label{a6}
\ln\det\frac{A_1}{A_0}=-\int_0^{\infty}\frac{d\tau}{\tau}\,\mathrm{tr}
\left(e^{-A_1\tau}-e^{-A_0\tau}\right),
\end{equation}
where the integral operators $e^{-A_1\tau}$ and $e^{-A_0\tau}$ are respectively associated with $K_1(x,y;\tau)$ and $K_0(x,y;\tau)$.

\section{Classical solution}
Let an $N$-dimensional vector $f(x)$ and an $N\times N$ matrix $B_{\mu}(x)$
with smooth coefficients be given. We calculate the operator $A_1$ as
\begin{equation}
\label{b00}
A_1f(x)=-D_{x^{\mu}}D_{x_{\mu}}f(x),
\end{equation}
where
\begin{equation}
\label{b0}
D_{x^{\mu}}f(x)=
(\partial_{x^{\mu}}+B_{\mu}(x))f(x).
\end{equation}
With regard to problem statement (\ref{a4}), we seek the solution in the form
\begin{equation}
\label{b2}
K_1(x,y;\tau)=K_0(x,y;\tau)\sum\limits_{n=0}^{\infty}
\tau^na_n(x,y),
\end{equation}
where the function
\begin{equation}
\label{b3}
K_0(x,y;\tau)=\frac{1}{(4\pi\tau)^
{d/2}}\exp{\left({-\frac{|x-y|^2}{4\tau}}\right)}
\end{equation}
is a solution of problem (\ref{a2}). We then have the following assertion.
\begin{lemma} 
\label{lem1}
Function (\ref{b2}) is a formal solution of problem (\ref{a2}) if
\begin{equation}
\label{111}
\begin{tabular}{c}
$(x-y)^{\mu}D_{x^{\mu}}a_0(x,y)=0,$\\
$\left((n+1)+
(x-y)^{\mu}D_{x^{\mu}}\right)a_{n+1}(x,y)=-A_1a_n(x,y),\,\,n\geqslant0.$
\end{tabular}
\end{equation}
\end{lemma}
\noindent Let the arguments $x$ and $y$ belong to $\mathbb{R}^d$. Then the operator $x^{\mu}\partial_{\mu}$ 
is the degree operator. Hence, if we assume that the solution can be expanded in a Taylor series in a neighborhood of a point, then each monomial of the form
\begin{equation*}
x_1^{\alpha_1}\cdot\ldots\cdot x_d^{\alpha_d},
\,\,\alpha_j\in\mathbb{N}\cup{0},\,\,\forall j\in\{1\ldots,d\},
\end{equation*}
is subject to the condition $\alpha_1+\ldots+\alpha_d=0$.
This immediately implies the alternative that either $\alpha_j=0$ for all $j\in\{1,\ldots,d\}$ or there exists a number $j\in{1,\ldots,d}$ such that
$\alpha_j<0$. A solution that can be expanded in a Taylor series at all points of the considered domain is said to be classical.

We introduce the operator of the P-ordered exponential for a fixed field $-B_{\mu}(x)$ and a contour $\gamma$ that is the segment connecting the initial and final points.
\begin{definition}
\label{def1}
The P-exponential is defined by the formula
\begin{equation}
\label{b18}
\Phi(x,y)=1+\sum\limits_{n=1}^{\infty}(-1)^n
\int_{0}^{1}\ldots\int_{0}^{1}
ds_1\ldots ds_n\,
\frac{dz_1^{\nu_1}}{ds_1}\ldots\frac{dz_n^{\nu_n}}{ds_n}
B_{\nu_1}(z_1)\ldots B_{\nu_n}(z_n),
\end{equation}
where the points $z_1,\ldots,z_{n-1}$ are located on the contour in the indexed order, the parameterizations $z_j(\cdot)$
are maps of the interval $[0,1]$ into a part of the interval from $y$ to $z_{j-1}$ for $j=2,\ldots,n$ and from $y$ to $x$ for $j=1$.
\end{definition}
The P-ordered exponential has the following basic properties:
\begin{itemize}
\item We have the equality
\begin{equation}
\label{nek2}
\Phi(x,x)=1.
\end{equation}
 \item Function (\ref{b18}) is a solution of the integral equation
\begin{equation}
\label{b19}
\Phi(x,y)=1-
\int_{0}^{1}
ds\,\frac{dz^{\nu}(s)}{ds}B_{\nu}(z(s))\,\Phi(z(s),y).
\end{equation}
\item The inverse operator has the form (see the proof in the appendix)
\begin{equation}
\label{nek3}
\Phi^{-1}(x,y)=\Phi(y,x).
\end{equation}
\end{itemize}
\begin{lemma}
\label{lem2}
The function $\Phi(x,y)$ is a classical solution in $\mathbb{R}^d$ of the problem
\begin{equation}
\label{b5}
(x-y)^{\mu}D_{x_{\mu}}a_0(x,y)=0,\,\,\,\,\,\,
\left.a_0(x,y)\right|_{x=y}=1.
\end{equation}
\end{lemma}
\noindent\textbf{Proof:}
It suffices to note that by choosing the parameterization in the form
$z^{\mu}(s)=(1-s)y^{\mu}+sx^{\mu}$
and integrating by parts, we can represent the derivative as
\begin{equation}
\label{b23}
\partial_{x^{\mu}}\left(\Phi(x,y)\right)=
-B_{\mu}(x)\Phi(x,y)-
\int_{0}^{1}
dz^{\nu}(s)\,sH_{\mu\nu}(z(s),y),
\end{equation}
where the operator $H_{\mu\nu}$ antisymmetric in the indices is given by
\begin{equation}
\label{b22}
H_{\mu\nu}(z,y)=\partial_{z^{\mu}}(
B_{\nu}(z)\Phi(z,y))-\partial_{z^{\nu}}(
B_{\mu}(z)\Phi(z,y)).
\end{equation}
$\blacksquare$
\begin{lemma}
\label{lem3}
If the three points $x$, $y$ and $z$ are collinear, then
\begin{equation}
\label{nek4}
\Phi(x,y)=\Phi(x,z)\Phi(z,y).
\end{equation}
\end{lemma}
\noindent\textbf{Proof:}
Because of the preceding properties and the vector collinearity condition
$(x-y)^{\mu}=\text{const}\cdot(x-z)^{\mu}$, the left- and right-hand sides of (\ref{nek4}) solve the problem
\begin{equation}
\label{nek5}
(x-y)^{\mu}(\partial_{x_{\mu}}+B_{\mu}(x))\Psi(x,y)=0,\,\,\,\,\,\,
\left.\Psi(x,y)\right|_{x=y}=1,
\end{equation}
which proves the proposition. $\blacksquare$

To solve the other differential equations in (\ref{111}), 
it suffices to understand how the solution of the following model problem is constructed. Let a sufficiently smooth and well-decreasing function $V(x,y)$
be given and its expansion in a Taylor series around a point $y$ have the form
\begin{equation*}
V(x,y)=\sum\limits_{k=0}^{\infty}
\frac{(x-y)_{\nu_1\ldots\nu_k}}{k!}V_k^{\nu_1\ldots\nu_k}(y)\,,
\end{equation*}
where we write
$(x-y)_{\nu_1\ldots\nu_n}:=(x-y)_{\nu_1}\cdot\ldots\cdot(x-y)_{\nu_n}$ for convenience. 
Then we have the following assertion.
\begin{lemma}
\label{lem4}
The classical solution in $\mathbb{R}^d$ of the equation
\begin{equation}
\label{b25}
\left((n+1)+
(x-y)^{\mu}D_{x^{\mu}}\right)W_{n+1}(x,y)=V(x,y)
\end{equation}
has the form
\begin{equation}
\label{b35}
W_{n+1}(x,y)=
\int_{0}^{1}ds\,
s^{n}
\Phi(x,z(s))V(z(s),y).
\end{equation}
\end{lemma}
\noindent\textbf{Proof:} It suffices to make the change
\begin{equation}
\label{b26}
W_{n+1}(x,y)=
e^{-(n+1)\pi_{\mu}
\ln{(x-y)^{\mu}}}\Phi(x,y)\widetilde{W}_{n+1}(x,y),
\end{equation}
where
\begin{equation*}
\pi_{\mu}\ln{(x-y)^{\mu}}=\sum\limits_{k=1}^{d}\pi_k\ln{(x-y)^{k}},\,\,
\pi\in\mathbb{R}^d:\,\,\sum\limits_{i=1}^{d}\pi_i=1.
\end{equation*}
We can then write the solution as
\begin{equation}
\label{b32}
\widetilde{W}_{n+1}(x,y)=
\text{const}(y)+\int_{0}^{1}ds\,s^{-1}
e^{(n+1)\pi_{\mu}\ln{(z(s)-y)^{\mu}}}
\Phi^{-1}(z(s),y)V(z(s),y).
\end{equation}
Solution (\ref{b35}) follows from the condition that the solution is bounded, Proposition \ref{lem2} and the expression
\begin{equation*}
e^{(n+1)\pi_{\mu}\ln{(z(s)-y)^{\mu}}}=
s^{n+1}e^{(n+1)\pi_{\mu}\ln{(x-y)^{\mu}}}.
\end{equation*}
$\blacksquare$
\begin{corollary}
It follows from Propositions \ref{lem1}--\ref{lem3} that the recursive system of classical solutions of differential equations (\ref{111}) with the initial condition has the form
\begin{equation}
\label{b36}
a_{n+1}(x,y)=-
\int_{0}^{1}ds\,
s^{n}
\Phi(x,z(s))(A_1a_n(z(s),y)).
\end{equation}
\end{corollary}

\section{Differentiation formulas for the P-exponential}
The formulas for differentiating the ordered exponential with the first and second arguments play a key role in constructing the diagram technique. In the case of covariant differentiation with respect to the first argument, a similar derivation can be found in
\cite{7}. We can prove the following assertion.
\begin{lemma}
\label{lem5} Let $x,y\in\mathbb{R}^d$ then we have the equality
\begin{equation}
\label{sho1}
D_{x^{\mu}}\Phi(x,y)=\int_{0}^{1}ds\,s\frac{dz^{\nu}}{ds}\Phi(x,z)
F_{\nu\mu}(z)\Phi(z,y),
\end{equation}
where
\begin{equation}
\label{sho2}
F_{\nu\mu}(x)=\partial_{x^{\nu}}B_{\mu}(x)-
\partial_{x^{\mu}}B_{\nu}(x)+[B_{\nu}(x),B_{\mu}(x)]
\end{equation}
and $s$ is the parameterization parameter, $z_{\nu}(s)=(1-s)y_{\nu}+sx_{\nu}$.
\end{lemma}
\noindent\textbf{Proof:} If we introduce the field
\begin{equation*}
K_{\mu}(y)=-\int_0^1dz^{\nu}(s)\,sH_{\mu\nu}(y,z(s))
\end{equation*}
and take relations (\ref{b23}) and (\ref{b22}) into account, then we obviously obtain
\begin{equation}
D_{x^{\mu}}\Phi(x,y)=K_{\mu}(x),\,\,\,\,\,\,(x-y)^{\nu}
K_{\nu}(z)\equiv0.
\end{equation}
We can also show that
\begin{equation}
\label{sho6}
K_{\mu}(x)=
-\int_0^1ds\,s(x-y)^{\nu}(F_{\mu\nu}\Phi(z,y)+
B_{\nu}(z)K_{\mu}(z)).
\end{equation}
Further, directly substituting (\ref{sho1}) in (\ref{sho6}) and taking property (\ref{b19}) into account, we see that the formula
to be proved holds. $\blacksquare$

The formula for differentiating the ordered exponential with respect to the second argument can be derived using (\ref{sho1}).
Indeed, if we differentiate the equality $\Phi^{-1}(x,y)\Phi(x,y)=1$ and take basic properties (\ref{nek2}), 
(\ref{nek3}) and (\ref{nek4}) into account, then we obtain the following assertion.
\begin{lemma}
\label{lem6}
Let $x,y\in\mathbb{R}^d$, then we have the equality
\begin{equation}
\label{dal1}
\partial_{x^{\mu}}\Phi(y,x)=\Phi(y,x)B_{\mu}(x)-\int_0^1ds\,s
\frac{dz^{\nu}}{ds}
\Phi(y,z)F_{\nu\mu}(z)\Phi(z,x).
\end{equation}
\end{lemma}
\noindent Formula (\ref{dal1}) is transformed after the changes $x\leftrightarrow y$ and
$s\rightarrow 1-s$, and we obtain the following assertion.
\begin{lemma}
\label{lem7}
Let $x,y\in\mathbb{R}^d$, then we have the equality
\begin{equation}
\label{dal4}
\partial_{y^{\mu}}\Phi(x,y)=\Phi(x,y)B_{\mu}(y)+
\int_0^1ds\,(1-s)\frac{dz^{\nu}}{ds}
\Phi(x,z)F_{\nu\mu}(z)\Phi(z,y).
\end{equation}
\end{lemma}

\section{Diagram technique}
\subsection{Motivation}
We note that the diagram technique presented here is not related to Feynman diagrams but is mainly used to obtain compact expressions and conveniently calculate and analyze the properties.\\

\noindent As already noted, a key role in constructing the diagram technique is played by formulas (\ref{sho1}), 
(\ref{dal1}) and (\ref{dal4}). Indeed, we note that under the covariant differentiation, the function $\Phi(x,y)$becomes an integral of the product of ordered exponentials and the field strength. Therefore, continuing the differentiation, we obtain only ordered exponentials and derivatives of the field strength. A more detailed analysis confirms this assumption and allows obtaining a method for controlling the coefficients arising under the action of the differentiation operators.
\subsection{Basic definitions}
The diagram technique consists of several basic elements.
\begin{definition} 
\label{def2}
A function $\Phi(x,y)$ is associated with a line with the arguments
\begin{figure}[H]
\centerline{\includegraphics[width=0.3\linewidth]{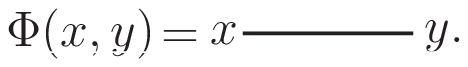}}
\end{figure}
\end{definition}
\begin{definition}
\label{def3}
A function between ordered exponentials is associated with a circle depending on the following parameters:
\begin{enumerate}
 \item a set of Greek indices $\mu_1\ldots\mu_n$ associated with
$$\nabla_{\mu_1}\ldots\nabla_{\mu_{n-1}}(d(z-y)^{\rho}F_{\rho\mu_n}(z)),$$
where all operators act on the variable $z$ and
$\nabla_{x_{\mu}}\cdot=\partial_{x^{\mu}}\cdot+[B_{\mu}(x)\,,\,\cdot\,]$,
 \item the parameterization parameter $s^k$ raised to an appropriate power, and
 \item a parameter  $(y\rightarrow x)$ representing the integral over the straight line from $y$ to $x$ with the parameterization $z_{\nu}=(1-s)y_{\nu}+sx_{\nu}$.
\end{enumerate}
\end{definition}
\noindent For example, formula (\ref{sho1}) becomes
\begin{figure}[H]
\centerline{\includegraphics[width=0.7\linewidth]{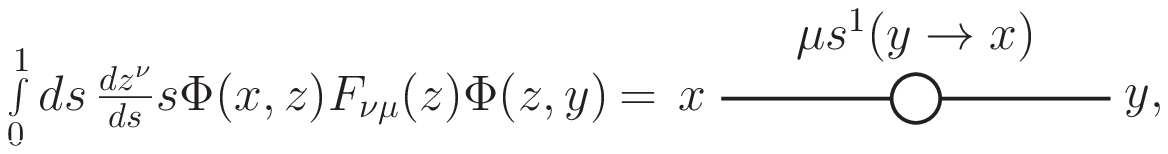}}
\end{figure}
\noindent where the two lines correspond to the functions $\Phi(x,z)$ and $\Phi(z,y)$ and the circle with the parameters $\mu$, $(y\rightarrow x)$, and $s^1$ corresponds to the integral from $x$ to $y$ of the form $dz_{\nu}(s)F_{\nu\mu}(z(s))$ with weight $s$. In the integration, the circle runs through the abovementioned interval.
\begin{definition}
\label{def4}
Let $z_{\nu}=(1-s)y_{\nu}+sx_{\nu}$. Then a construction of the form
\begin{figure}[H]
\centerline{\includegraphics[width=0.5\linewidth]{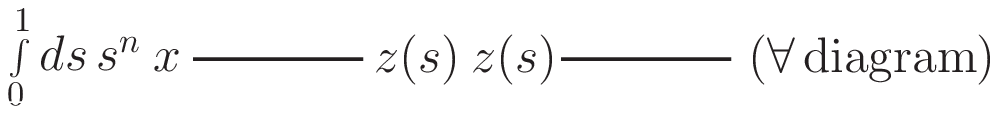}}
\end{figure}
corresponds to the diagram
\begin{figure}[H]
\centerline{\includegraphics[width=0.3\linewidth]{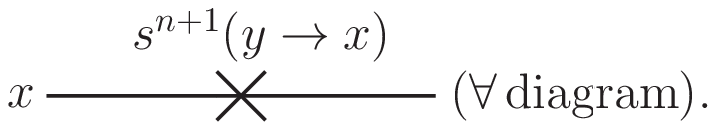}}
\end{figure}
\end{definition}
\noindent This definition is based on formula (\ref{b36}).

\subsection{Main example}
To understand the structure of the diagram technique, it is expedient to cal- culate the first iteration of system (\ref{b36}). 
In the zeroth order, the solution of system (\ref{111}) s determined by the formula $a_0(x,y)=\Phi(x,y)$ in Definition \ref{def1} (see formula (\ref{b18})). Further, we apply the Laplace operator and the integration operator.

First, by formula (\ref{sho1}), we have
\begin{equation}
\label{pri1}
D_{x^{\mu}}\Phi(x,y)=\int\limits_{0}^{1}ds_1\,s_1\frac{dz^{\nu}}{ds_1}\Phi(x,z)F_{\nu\mu}(z)\Phi(z,y),
\end{equation}
или
\begin{figure}[H]
\centerline{\includegraphics[width=0.5\linewidth]{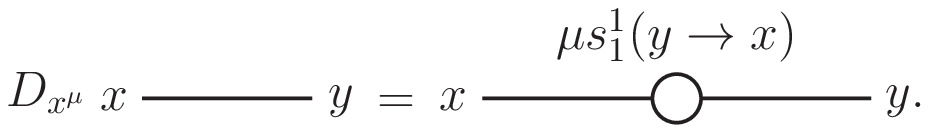}}
\end{figure}
Second, it follows from relations (\ref{sho1}) and (\ref{dal4}) that the second covariant derivative has the form
\begin{multline}
\label{pri2}
D_{x^{\mu}}D_{x^{\mu}}\Phi(x,y)=
\int_{y}^{x}dz^{\nu}(s_1)\,s_1\int_{z}^{x}
dz'^{\rho}(s_2)\,s_2\Phi(x,z')
F_{\rho\mu}(z')\Phi(z',z)
F_{\nu\mu}(z)\Phi(z,y)+\\+
\int_{y}^{x}dz^{\nu}(s_1)\,s_1^2
\int_z^xdz'^{\rho}(s_2)\,(1-s_2)
\Phi(x,z')F_{\rho\mu}(z')\Phi(z',z)
F_{\nu\mu}(z)\Phi(z,y)+\\+\int_{y}^{x}dz^{\nu}(s_1)\,s_1^2\Phi(x,z)F_{\nu\mu}(z)
\int_{y}^{z}dz'^{\rho}(s_2)\,s_2\Phi(z,z')
F_{\rho\mu}(z')\Phi(z',y)+\\+
\int_{y}^{x}s_1
\Phi(x,z)\left(s_1\partial_{z^{\mu}}
(dz^{\nu}(s_1)F_{\nu\mu}(z))\right)\Phi(z,y)+
\int_{y}^{x}dz^{\nu}(s_1)\,s_1^2\Phi(x,z)B^{\mu}(z)
F_{\nu\mu}(z)\Phi(z,y)-\\-
\int_{y}^{x}dz^{\nu}(s_1)\,s_1^2\Phi(x,z)F_{\nu\mu}(z)B^{\mu}(z)
\Phi(z,y),
\end{multline}
which in the diagram language is equivalent to
\begin{figure}[H]
\centerline{\includegraphics[width=0.8\linewidth]{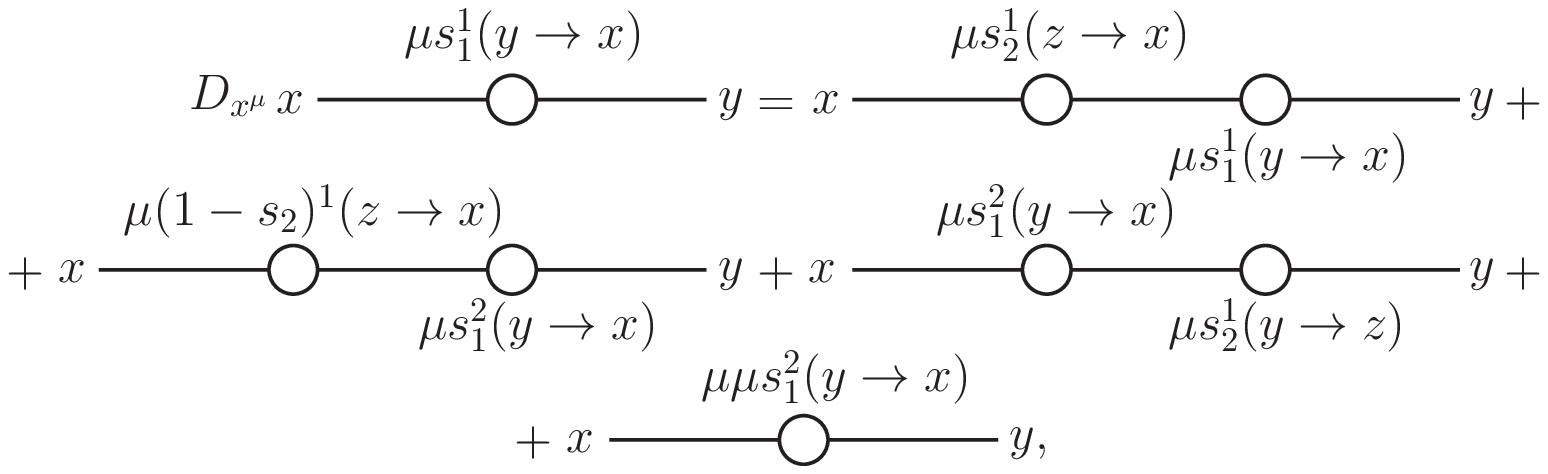}}
\end{figure}
\noindent where the first three terms in (\ref{pri2}) correspond to the first three diagrams and the last three terms form the
action of the operator $\nabla_{\mu}$ on the field strength and are depicted by the fourth diagram.
Third, after integration, we obtain
\begin{figure}[H]
\centerline{\includegraphics[width=0.8\linewidth]{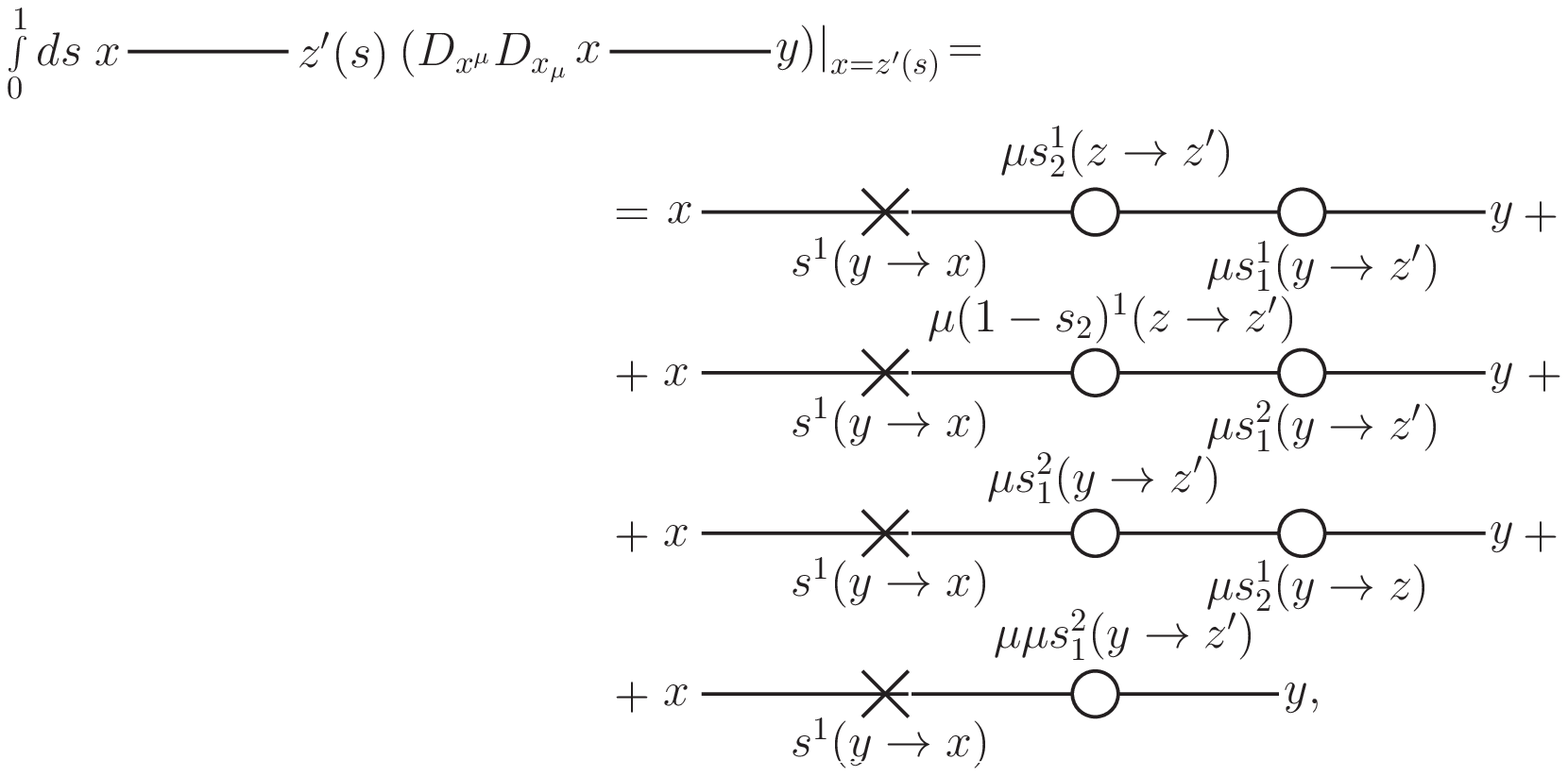}}
\end{figure}
\noindent where we use the standard parameterization $z_{\mu}'(s)=(1-s)y_{\mu}+sx_{\mu}$ and Definition \ref{def4}.

\subsection{Theorem on the differentiation of diagrams}
The main problem in this section is to learn how the covariant derivative can be applied to an arbitrary diagram, to simplify the diagram technique, and to eliminate unnecessary parameters.
\begin{lemma}
\label{lem8}
Let two continuous functions $f(x)$ and $g(x)$ and the parameterization
\begin{equation}
\label{teo1}
z_{\mu}(s)=(1-s)y_{\mu}+sy'_{\mu} ,\,\,\,\,\,\,x_{\mu}(t)=(1-t)z_{\mu}+ty'_{\mu}
\end{equation}
be given. Then
\begin{equation}
\label{teo2}
\int_{y}^{y'}dz_{\rho}\int_{z}^{y'}dx_{\nu}\,s^nt^pf(z)g(x)=
\int_{y}^{y'}dx'_{\nu}\int_{y}^{x'}dz'_{\rho}\,(s')^{n}(t')^{n+p}\left(\frac{1-s'}{1-s't'}\right)^pf(z')g(x'),
\end{equation}
where
\begin{equation}
\label{teo3}
x'_{\mu}(t')=(1-t')y_{\mu}+t'y'_{\mu} ,\,\,\,\,\,\,z'_{\mu}(s')=(1-s')y_{\mu}+s'x'_{\mu}.
\end{equation}
\end{lemma}
\noindent\textbf{Proof:} It suffices to change
\begin{equation}
t\longrightarrow t'=(1-s)t+s, 
\end{equation}
then to use the Fubini theorem
\begin{equation}
(s,t')\in[0,1]\times[s,1]\longrightarrow(s,t')\in[0,t']\times[0,1],
\end{equation}
and to make another change $s\rightarrow s/t'$.$\blacksquare$
\begin{lemma}
\label{lem9}
We have the equality
\begin{figure}[H]
\centerline{\includegraphics[width=0.7\linewidth]{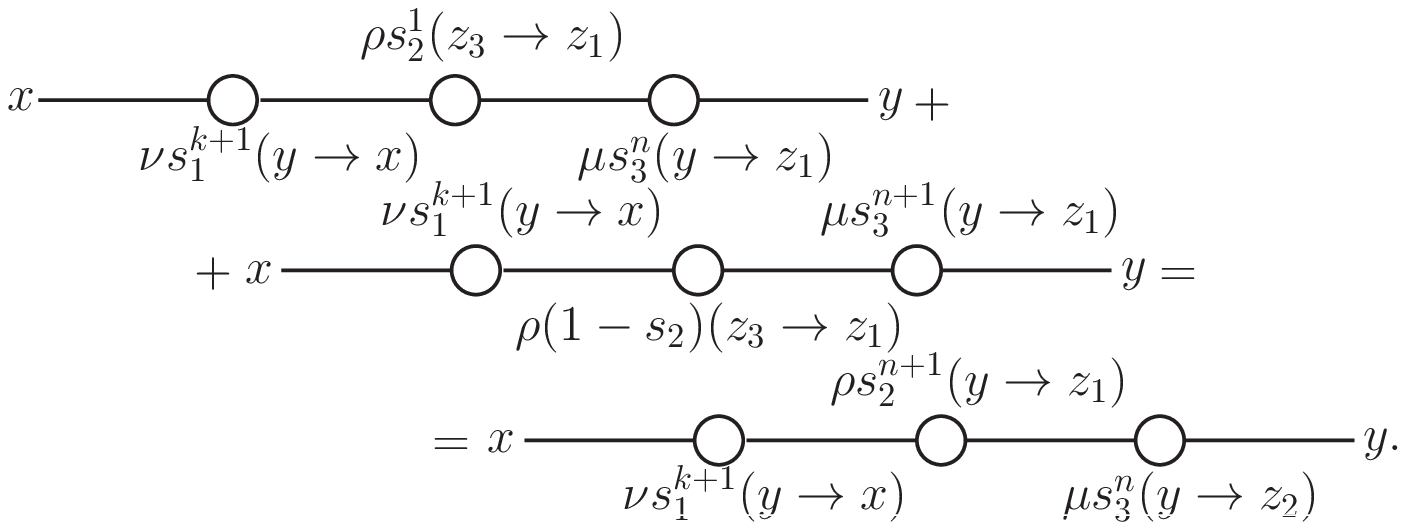}}
\end{figure}
\end{lemma}
\noindent\textbf{Proof:} This assertion can be proved in several stages. Proposition \ref{lem8} on the permutation of integrals implies that
\begin{figure}[H]
\centerline{\includegraphics[width=0.7\linewidth]{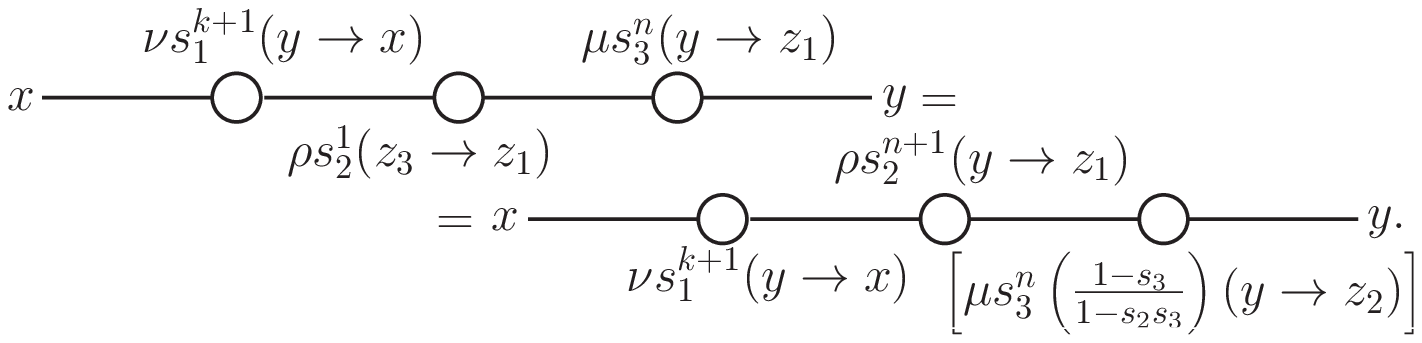}}
\end{figure}
\noindent Proposition \ref{lem8} similarly implies the result for the second diagram
\begin{figure}[H]
\centerline{\includegraphics[width=0.7\linewidth]{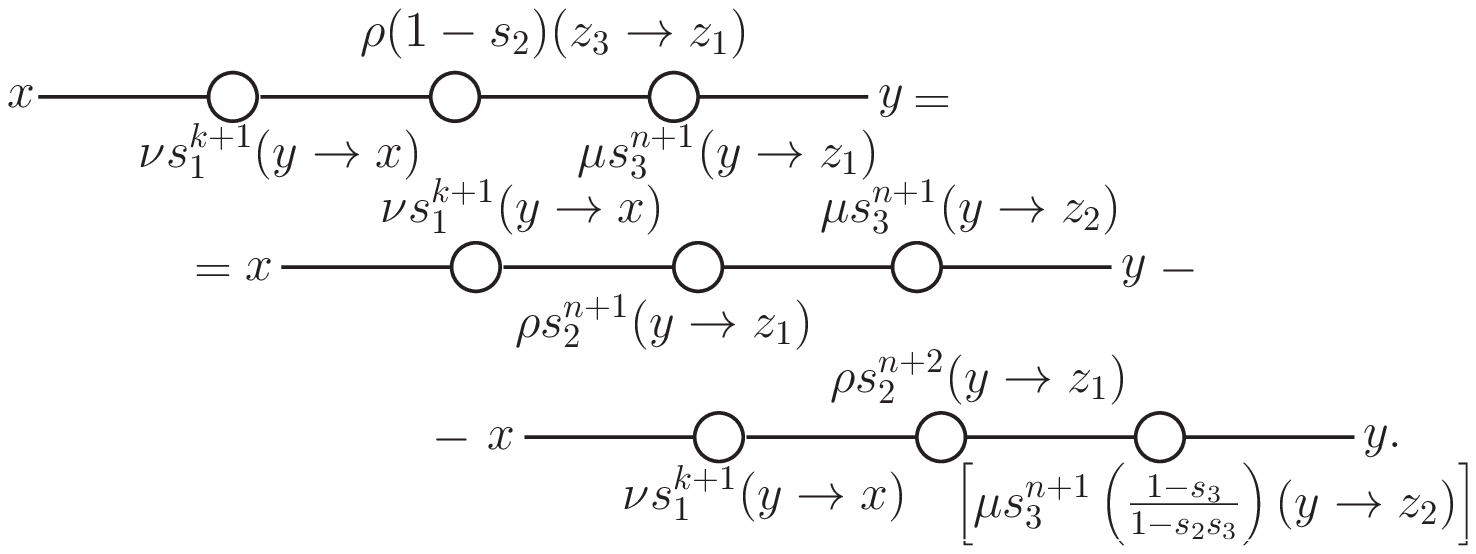}}
\end{figure}
\noindent The parameters satisfy the relations
\begin{equation}
\label{teo17}
s_2^{n+1}s_3^{n}\left(\frac{1-s_3}{1-s_2s_3}\right)+s_2^{n+1}s_3^{n+1}-s_2^{n+2}s_3^{n+1}\left(\frac{1-s_3}{1-s_2s_3}\right)=s_2^{n+1}s_3^{n}.
\end{equation}
$\blacksquare$
\begin{lemma}
\label{lem10}
We have the equality
\begin{figure}[H]
\centerline{\includegraphics[width=0.7\linewidth]{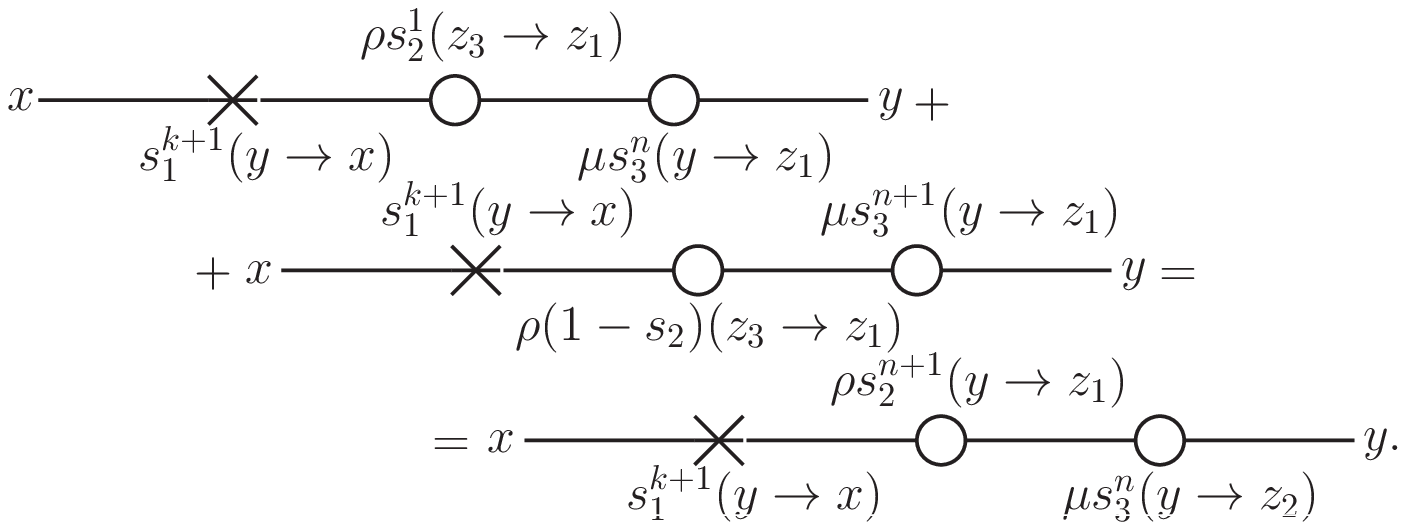}}
\end{figure}
\end{lemma}
\begin{lemma}
\label{lem11}
We have the equality
\begin{figure}[H]
\centerline{\includegraphics[width=0.7\linewidth]{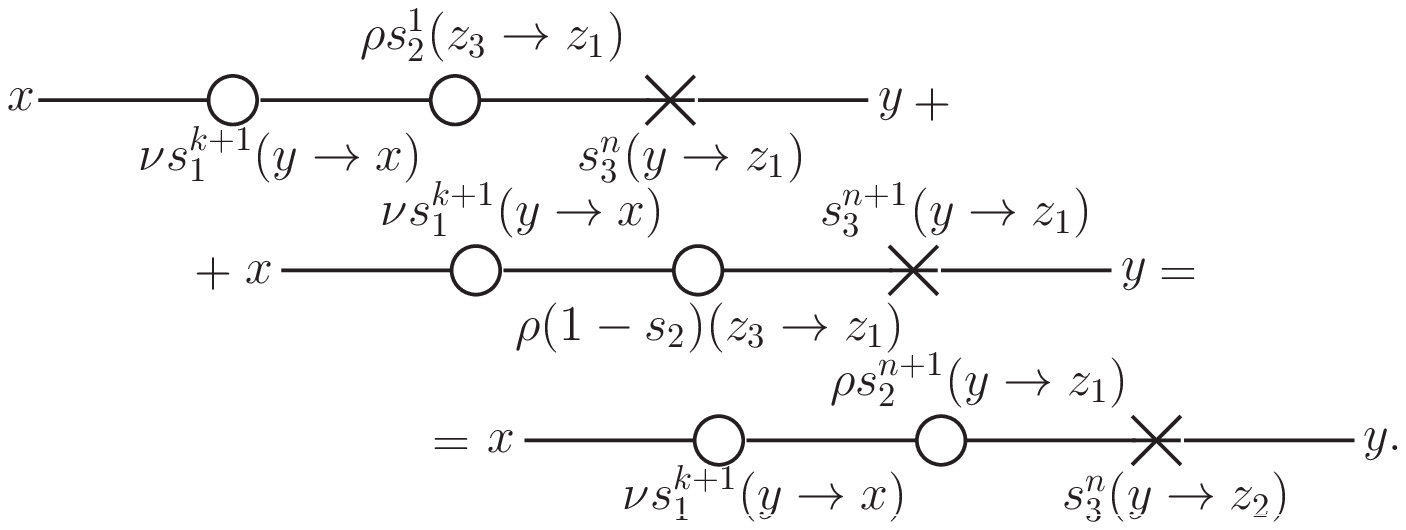}}
\end{figure}
\end{lemma}
\begin{corollary}
It follows from Propositions \ref{lem9}--\ref{lem11} that we can neglect the third parameter in Definition \ref{def3} because the integrals are always taken from the point $y$ to the nearest point on the left.
\end{corollary}
\begin{corollary} The formula for the first order becomes
\begin{figure}[H]
\centerline{\includegraphics[width=0.8\linewidth]{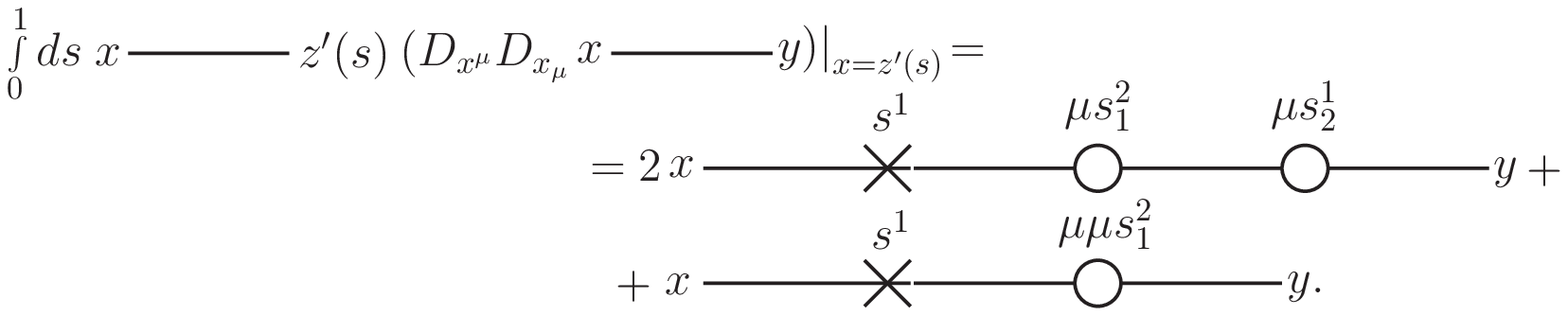}}
\end{figure}
\end{corollary}
\noindent Summarizing, we formulate a theorem on the differentiation of diagrams.
\begin{theorem}
\label{th1}
Let there be an arbitrary diagram with $i$ lines, $k$ circles, and $j$ crosses and the operation $D_{\rho}$ be applied to the diagram. Let
$t$ be a new parameterization. Then the following three cases are possible:
\begin{enumerate}
 \item If the derivative acts on a line, then the line is replaced with a circle with two lines on the sides and with the parameters $\rho$ and $t$ raised to the power equal to the power of the second parameter on the vertex to the right incremented by 1 (if the line is the rightmost, then the power is equal to 1). In this case, the power of each weight to the left of the differentiated line increases by 1. As a result, we have $i$ diagrams.
 \item If the derivative acts on a circle, then the power of its second parameter increases by 1, and the index $\rho$ is attached on the left to the others. In this case, the power of each second parameter of the vertices to the left increases by 1. As a result, we have $k$ diagrams.
 \item If the derivative acts on a cross, then the diagram vanishes.
\end{enumerate}
\end{theorem}
\noindent We therefore have $i+k$ diagrams after differentiation.\\
\noindent For example,
\begin{figure}[H]
\centerline{\includegraphics[width=0.7\linewidth]{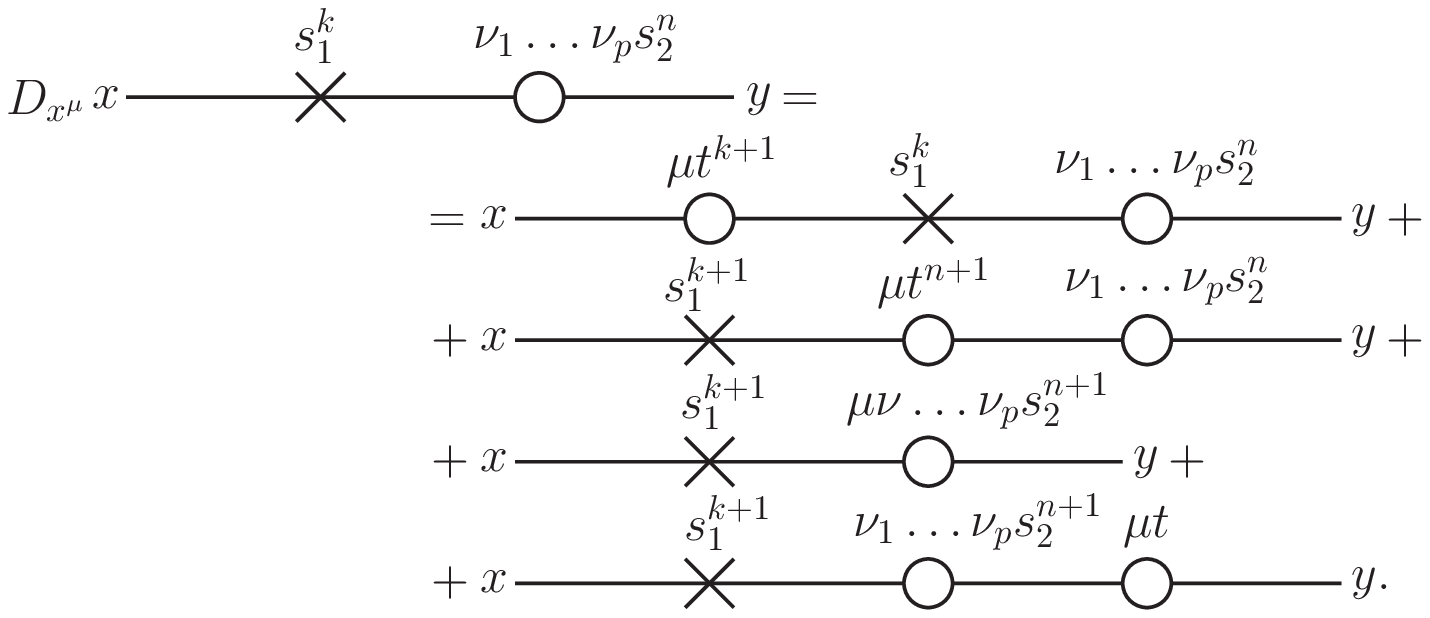}}
\end{figure}
\begin{corollary}
The differentiation operation splits into the operation $(\nabla)$ acting on each circle and the operation of addition of a new circle (complying with the rule of increase in the powers of the parameteri- zation parameters).
\end{corollary}

\subsection{First order}
The first order can be written as
\begin{figure}[H]
\centerline{\includegraphics[width=0.7\linewidth]{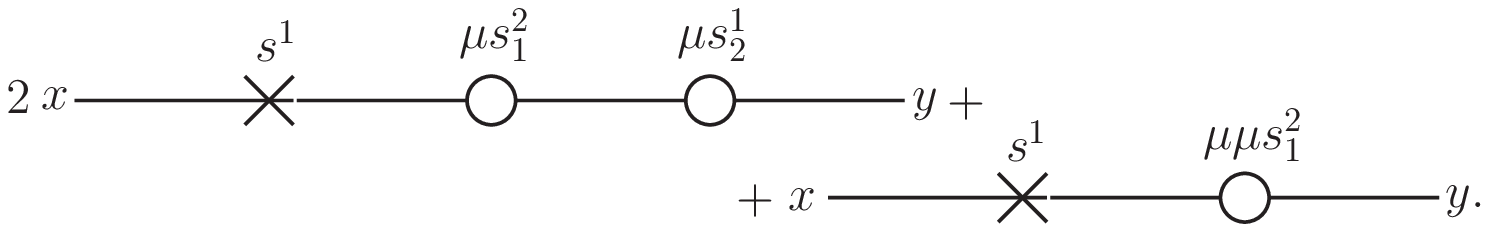}}
\end{figure}
\noindent To determine the contribution obtained by calculating the trace, we must take $y=x$.
Hence, we must determine the zeroth-order term of the Taylor series expansion. It is easy to see that the first term of the diagram starts from the quadratic term because each circle with a single index contains a linear contribution. In turn, the second term gives the result $\nabla^{\mu}((x-y)^{\nu}F_{\nu\mu})|_{x=y}$, which is obviously equal to zero because the field strength is antisymmetric.

\subsection{Second order}
To calculate the second order, it suffices to apply the covariant derivative twice to the first order and integrate. Further, we must set $y=x$ and choose the contributions with nonzero traces. We must take into account that, first, each circle must contain at least two Greek indices because the contribution for $y=x$ is otherwise zero; second, a circle with two equal Greek indices contributes zero; third, a diagram with one circle and with more than two Greek indices contributes zero. With these remarks taken into account, the formula becomes
\begin{figure}[H]
\centerline{\includegraphics[width=0.6\linewidth]{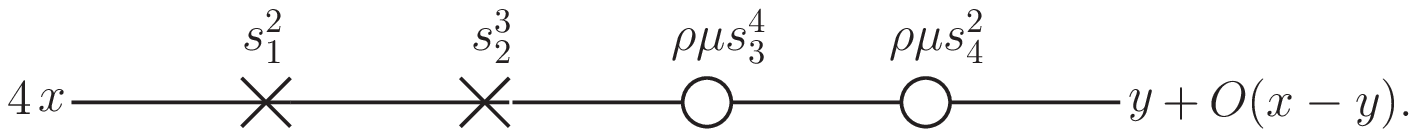}}
\end{figure}
\noindent For $y=x$, all ordered exponentials become the unit. Hence, using the relation
\begin{equation}
\label{vto1}
\left.\left(\nabla_{\rho}((x-y)^{\nu}F_{\nu\mu}(x))\right)\right|_{x=y}=F_{\rho\mu}(x),
\end{equation}
we can write the answer as
\begin{equation}
\label{vto2}
\frac{1}{12}F_{\rho\mu}(x)F^{\rho\mu}(x).
\end{equation}

\section{Operators acting on matrices}

\subsection{Motivation}
The rules of the diagram technique described in the preceding section easily allow determining the coefficients. As an example, we obtained standard results (\ref{vto2}) for the first and second orders. But the procedure is recursive. In this section, we present a formalism that allows deriving the final series for the heat kernel of the covariant Laplace operator for $y=x$.

\subsection{Matrices}
It was previously shown that the diagram vertices have two parameters and each diagram can therefore be associated with a matrix with two rows. The number of columns in such a matrix is equal to the total number of circles and crosses in the diagram.

The matrices have the following basic properties:
\begin{enumerate}
 \item The matrices have two rows and $n$ columns preserving the order, where $n$ is the total number of vertices in the diagram.
 \item The elements in the first row can be of two types: either $\mathbf{1}$, 
 if the column corresponds to a cross in the diagram, or a set of Greek indices, if the column corresponds to a circle.
 \item The second row contains powers of the second parameters of the vertices.
 \item The unit \(\mathbb{1}\) denotes an element that does not contain columns.
\end{enumerate}
We thus obtain matrices of the form
$\mathbb{1}, \text{M}^{2\times1}, \text{M}^{2\times2},\ldots.$ We note that in each order of the heat kernel, there are finitely many matrices of a finite size. Such matrices can be added only if they have the same elements and the same size. It is clear that the basis in this case is at most countable because the first and the second row can contain only elements from a countable set.

\subsection{Definitions of the operators}
According to the rules listed above, any diagram can be written as a matrix, and it is hence convenient to introduce an operator calculating the diagram at the point $y=x$.\\

\begin{definition}
\label{def5} An operator
$\Upsilon$ associates each set of Greek indices with a structure corresponding to it by Definition \ref{def3}, where the differential $d(z-y)$ is replaced with $z-y$, the point $z=y=x$ is chosen, and the product of elements of the upper row is then divided by the product of elements of the lower row.
\end{definition}
For example,
\begin{equation}
\label{oper00}
\Upsilon\mathbb{1}=1,\,\,\,\,\,\,
\Upsilon
\begin{pmatrix}
1&\nu\mu\\
2 & 3 
\end{pmatrix}
=\frac{1}{6}(\nabla_{z^{\nu}}((z-y)^{\rho}F_{\rho\mu}(z)))|_{z=y=x}.
\end{equation}
It was previously noted that the covariant derivative of the diagram reduces either to the operator of addition of a circle with an index or to the operator of index addition (complying with the rule of variation in the weight powers). The integration operator also reduces to the operator of cross addition. It is easy to see that we can introduce the following similar operators acting on matrices:
\begin{definition}
\label{def6}
The operator of additional multiplication
\begin{equation}
\label{oper1}
B^{\mu}
\begin{pmatrix}
\nu_1&\nu_2 & \ldots & \nu_n\\
k_1 & k_2 & \ldots & k_n
\end{pmatrix}
=
\begin{pmatrix}
\mu\nu_1&\nu_2 & \ldots & \nu_n\\
k_1+1 & k_2 & \ldots & k_n
\end{pmatrix}
+
\ldots+
\begin{pmatrix}
\nu_1&\nu_2 & \ldots & \mu\nu_n\\
k_1+1 & k_2+1 & \ldots & k_n+1
\end{pmatrix}.
\end{equation}
\end{definition}
\begin{definition}
\label{def7}
The operator of column addition
\begin{multline}
\label{oper2}
A^{\mu}
\begin{pmatrix}
\nu_1&\nu_2 & \ldots & \nu_n\\
k_1 & k_2 & \ldots & k_n
\end{pmatrix}
=
\begin{pmatrix}
\mu & \nu_1&\nu_2 & \ldots & \nu_n\\
k_1+1 & k_1 & k_2 & \ldots & k_n
\end{pmatrix}
+\ldots\\
\ldots+
\begin{pmatrix}
\nu_1&\nu_2 & \ldots & \mu & \nu_n\\
k_1+1 & k_2+1 & \ldots & k_n+1 &k_n
\end{pmatrix}
+
\begin{pmatrix}
\nu_1&\nu_2 & \ldots & \nu_n & \mu\\
k_1+1 & k_2+1 & \ldots & k_n+1 & 1
\end{pmatrix}.
\end{multline}
\end{definition}
\begin{definition}
\label{def8}
The integration operator
\begin{equation}
\label{oper3}
S_l^{\mu/\mathbf{1}}
\begin{pmatrix}
\nu_1&\nu_2 & \ldots & \nu_n\\
k_1 & k_2 & \ldots & k_n
\end{pmatrix}
=
\begin{pmatrix}
\mu/\mathbf{1} & \nu_1&\nu_2 & \ldots & \nu_n\\
l & k_1 & k_2 & \ldots & k_n
\end{pmatrix}.
\end{equation}
\end{definition}
\noindent Using these definitions, we can write heat kernel (\ref{b2}) for the covariant Laplace operator in the form
\begin{equation}
\label{rt1}
K_1(x,x;\tau)=\Upsilon\left(1+
\sum\limits_{n=1}^{\infty}\tau^{n}
\prod_{k=1}^{n}S^{\mathbf{1}}_k(A^{\mu_k}+B^{\mu_k})(A^{\mu_k}+B^{\mu_k})\right)
\mathbb{1}.
\end{equation}

\subsection{Commutators of the operators}
The main idea is to use the commutators of the operators
defined above to move the operators of additional multiplication and of column addition to the right in (\ref{rt1}) and thus avoid the recursive procedure. The following assertion can easily be verified.
\begin{lemma}
\label{rt2}
We have the equalities
\begin{equation}
\label{111}
\begin{tabular}{ll}
$A^{\mu}S^{\mathbf{1}}_k=S^{\mathbf{1}}_{k+1}A^{\mu}+S^{\mu}_{k+1}S^{\mathbf{1}}_k,$& 
$A^{\mu}S^{\nu_1\ldots\nu_n}_k=
S^{\nu_1\ldots\nu_n}_{k+1}A^{\mu}+S^{\mu}_{k+1}S^{\nu_1\ldots\nu_n}_k,$\\
$B^{\mu}S^{\mathbf{1}}_k=S^{\mathbf{1}}_{k+1}B^{\mu},$&
$B^{\mu}S^{\nu_1\ldots\nu_n}_k=
S^{\nu_1\ldots\nu_n}_{k+1}B^{\mu}+S^{\mu
\nu_1\ldots\nu_n}_{k+1}.$
\end{tabular}
\end{equation}
\end{lemma}
This assertion gives only the relations used in the further proofs.

\subsection{Properties of the integration operators}
Relations (\ref{rt2}) show that the commutators of the operators $A$ and $B$ with the operators $S$ again become $S$ operators. 
The whole construction for the heat kernel is therefore reducible to a sum of combinations of integration operators. Hence, we must introduce several structures to simplify the further description.

First, we introduce the following operator series. Let $I_n=\{n,\ldots,1\}$ and let $\mu_{I_n}=\mu_n\ldots\mu_1$ be a multi-index. Then
\begin{equation}
\label{rt6}
\Sigma_l^{\mu_{I_n}}=\sum\limits_{\sigma}
S_{l+\sharp\,\cup_{i=1}^{k}\sigma_i}^{\mu_{\sigma_k}}\ldots
S_{l+\sharp \sigma_1\cup \sigma_2}^{\mu_{\sigma
_2}}S_{l+\sharp\,\sigma_1}^{\mu_{\sigma_1}},
\end{equation}
where the sums are taken over partitions with the properties
\begin{itemize}
 \item $\sigma_j\subset I_n$ for all $j\in \{1,\ldots,k\},$
 \item $\sigma_i\cap \sigma_j=\varnothing$ for $i\neq j,$ 
 \item $\cup_{i=1}^{k}\sigma_i=I_n,$ and
 \item $\sigma_s(i)>\sigma_s(j)$ for any $s\in\{1,\ldots,k\}$ and any $i,j\in\{1,\ldots,\sharp \sigma_s\}$ such that $i>j.$
\end{itemize}
Here, the operator $\sharp$ counts the number of elements in a set. The sum in formula (\ref{rt6}) is finite, and there are hence no problems related to convergence.

If we act by such an operator on $\mathbb{1}$ (this does not impose any restrictions, because any matrix can be obtained from the unit by a certain set of integration operators) and then apply the operator $\Upsilon$, then we consequently obtain a sum containing the field strengths and numerical coefficients. To obtain the exact result, we must take several assertions into account.
\begin{lemma}
\label{lem13}
We have the equalities
\begin{equation}
\label{rt7}
\nabla_{\mu_{I_n}}((x-y)^{\rho}F_{\rho\nu}(x))=(x-y)^{\rho}\nabla_{\mu_{I_n}}F_{\rho\nu}(x)+\sum\limits_{k=1}^{n}\nabla_{\mu_{I_n\setminus k}}F_{\mu_k\nu}(x),
\end{equation}
\begin{equation}
\label{rt8}
\left.\nabla_{\mu_{I_n}}((x-y)^{\rho}F_{\rho\nu}(x))\right|_{x=y}=\sum\limits_{k=1}^{n}\nabla_{\mu_{I_n\setminus k}}F_{\mu_k\nu}(x).
\end{equation}
\end{lemma}
\begin{lemma}
\label{lem14}
We have the equalities
\begin{equation}
\label{rt9}
\widehat{S}_l^{\mu_{I_n}\nu}=\Upsilon S_l^{\mu_{I_n}\nu}\mathbb{1}=\frac{1}{l}\sum\limits_{k=1}^{n}\nabla_{\mu_{I_n\setminus k}}F_{\mu_k\nu}(x);
\end{equation}
where we introduce the additional notation $\widehat{S}_l^{\mu_{I_n}\nu}$ for convenience.
\end{lemma}
\begin{lemma}
\label{lem15}
For any sets of indices $I$ and $J$ and positive integers $l$ and $k$, we have the relation
\begin{multline}
\label{rt10}
\Upsilon (S_l^{\mu_{I}\nu}S_k^{\mu_{J}\rho}\mathbb{1})=
(\Upsilon S_l^{\mu_{I}\nu}\mathbb{1})
(\Upsilon S_k^{\mu_{J}\rho}\mathbb{1})=\\=
\frac{1}{l\cdot k}\left(
\sum\limits_{i=1}^{\sharp I}
\nabla_{\mu_{I\setminus i}}F_{\mu_i\nu}(x)\right)\left(
\sum\limits_{j=1}^{\sharp J}
\nabla_{\mu_{J\setminus j}}F_{\mu_j\rho}(x)\right)=
\widehat{S}_l^{\mu_{I}\nu}
\widehat{S}_k^{\mu_{J}\rho}.
\end{multline}
\end{lemma}
\noindent The proof of this assertion follows from Propositions \ref{lem13} and \ref{lem14}. It is clear that this property can be generalized to arbitrarily many integration operators. As a result, we can state that the structure $\Upsilon \Sigma \mathbb{1}$ is completely known and can be expressed in terms of the field strength and its derivatives according to the following assertion.
\begin{lemma}
\label{lem16}
We have the equality
\begin{equation}
\label{rt11}
\Upsilon\Sigma_l^{\mu_{I_n}}\mathbb{1}=\sum\limits_{\sigma} 
\widehat{S}_{l+\sharp\,\cup_{i=1}^{k}\sigma_i}^{\mu_{\sigma_k}}\ldots
\widehat{S}_{l+\sharp \sigma_1\cup \sigma_2}^{\mu_{\sigma
_2}}\widehat{S}_{l+\sharp\,\sigma_1}^{\mu_{\sigma_1}},
\end{equation}
where the summation complies with the rules described above.
\end{lemma}
\noindent\textbf{Remark:} Proposition \ref{lem15} also holds in the case where the integration operators $S$ are replaced with the operators $\Sigma$. This assertion can also be generalized to the case of arbitrarily many factors.

\subsection{Series for the heat kernel}
Formula (\ref{rt1}) is a Taylor series in powers of the proper time. A separate Seeley–DeWitt coefficient has the form
\begin{equation}
\label{ra1}
a_n(x,x)=\Upsilon\left(
\prod_{k=1}^{n}S^{\mathbf{1}}_k(A^{\mu_k}+B^{\mu_k})(A^{\mu_k}+B^{\mu_k})\right)
\mathbb{1}.
\end{equation}
To write the coefficient in terms of the field strength and its derivatives, we need the following assertion.
\begin{lemma}
\label{lem17}
Let $I$ be a set of indices with $\sharp I=n$ and $l$ be a positive integer. Then we have the operator equality
\begin{equation}
\label{ra2}
(A+B)^{\mu_{I}}S_l^{\mathbf{1}}=\sum\limits_{k=0}^{n}\sum\limits_{\sigma_k}\Sigma_{l+k}^{\mu_{I\setminus\sigma_k}}S_{l+k}^{\mathbf{1}}(A+B)^{\mu_{\sigma_k}},
\end{equation}
where the sums are taken over all possible subsets $\sigma_k\subset I$ satisfying the conditions $\sharp\sigma_k=k$ and
$\sigma_k(i)>\sigma_k(j)$ for all $i,j\in\{1,\ldots,k\}$ such that $i>j$.
\end{lemma}
\noindent\textbf{Proof:} We verify formula (\ref{ra2}) by induction. For $n=1$, everything is obvious:
\begin{equation}
\label{ra3}
(A+B)^{\mu_1}S_l^{\mathbf{1}}=S_{l+1}^{\mathbf{1}}(A+B)^{\mu_1}+S_{l+1}^{\mu_1}S_l^{\mathbf{1}}.
\end{equation}
We suppose that (\ref{ra2}) is satisfied for $n=j$ and prove it for $n=j+1$:
\begin{multline}
\label{ra4}
(A+B)^{\rho}(A+B)^{\mu_{I}}S_l^{\mathbf{1}}=\sum\limits_{k=0}^{j}
\sum\limits_{\sigma_k}(A+B)^{\rho}
\Sigma_{l+k}^{\mu_{I\setminus\sigma_k}}
S_{l+k}^{\mathbf{1}}(A+B)^{\mu_{\sigma_k}}=\\=\sum\limits_{k=0}^{j}
\sum\limits_{\sigma_k}\left(
\Sigma_{l+k}^{\rho\mu_{I\setminus\sigma_k}}
S_{l+k}^{\mathbf{1}}(A+B)^{\mu_{\sigma_k}}+
\Sigma_{l+k+1}^{\mu_{I\setminus\sigma_k}}
S_{l+k+1}^{\mathbf{1}}(A+B)^{\rho\mu_{\sigma_k}}\right).
\end{multline}
It remains to redefine the indices, and this implies (\ref{ra2}). The proof of the proposition is complete. $\blacksquare$

It is more convenient to substitute formula (\ref{ra2}) in (\ref{ra1}) in several steps.\\
\begin{lemma}
\label{lem18} Let $I=I_{2n}$. Then
\begin{multline}
\label{ra5}
\prod_{k=1}^{n}S^{\mathbf{1}}_k(A^{\mu_{2k}}+B^{\mu_{2k}})
(A^{\mu_{2k-1}}+B^{\mu_{2k-1}})=\\=
\sum\limits_{p_1=0}^{2}
\sum\limits_{p_2=0}^{4-p_1}\ldots
\sum\limits_{p_{n-1}=0}^{2(n-1)-\sum_{k=1}^{n-2}p_{k}}
\sum\limits_{\sigma}
S_n^{\mathbf{1}}\Sigma_{n+1-p_1}^{\mu_{\sigma_{p_1}}}S_{n+1-p_1}^{\mathbf{1}}
\Sigma_{n+2-p_1-p_2}^{\mu_{\sigma_{p_2}}}S_{n+2-p_1-p_2}^{\mathbf{1}}\ldots\\\ldots
\Sigma_{2n-1-\sum_{j=1}^{n-1}p_{j-1}}^{\mu_{\sigma_{p_{n-1}}}}S_{2n-1-\sum_{j=1}^{n-1}p_{j-1}}^{\mathbf{1}}
(A+B)^{\mu_{I\setminus\bigcup_{s=1}^{n-1}\sigma_{p_s}}},
\end{multline}
where the sums are taken over all possible subsets $\sigma_{p_{k}}$ satisfying the conditions $\sharp\sigma_{p_k}=p_k$,
$\sigma_{p_{k}}\subset(I_{2n}\setminus I_{2(n-k)})\setminus
(\cup_{j=1}^{k-1}\sigma_{p_{j}})$, and $\sigma_{p_{k}}(i)>\sigma_{p_{k}}(j)$ for all $i,j\in\{1,\ldots,p_k\}$ such that $i>j$.
\end{lemma}
\noindent To prove (\ref{ra5}), we apply formula (\ref{ra2}) several times.
\begin{lemma}
\label{lem19}
Let $I$ be a set of indices. Then
\begin{equation}
\label{ra6}
\Upsilon(A+B)^{\mu_I}\mathbb{1}=\Upsilon\Sigma^{\mu_I}_0\mathbb{1}
=\widehat{\Sigma}^{\mu_I}_0.
\end{equation}
\end{lemma}
This assertion follows from the definition of operators. At the last stage, we must act by operator (\ref{ra5}) on the unit and then multiply from the right by the operator $\Upsilon$. We can formulate the final theorem summarizing the obtained results.
\begin{theorem}
\label{th2}
Let $\mu_{2k-1}=\mu_{2k}$ for all $k\in\{1,\ldots,n\}$. Then
\begin{multline}
\label{ra7}
a_n(x,x)=
\sum\limits_{p_1=0}^{2}
\sum\limits_{p_2=0}^{4-p_1}\ldots
\sum\limits_{p_{n-1}}^{2(n-1)-\sum_{k=1}^{n-2}p_{k}}
\sum\limits_{\sigma}
\frac{\widehat{\Sigma}_{n+1-p_1}^{\mu_{\sigma_{p_1}}}}{n+1-p_1}
\frac{\widehat{\Sigma}_{n+2-p_1-p_2}^{\mu_{\sigma_{p_2}}}}{n+2-p_1-p_2}
\ldots\\\ldots
\frac{\widehat{\Sigma}_{2n-1-\sum_{j=1}^{n-1}p_{j-1}}^{\mu_{\sigma_{p_{n-1}}}}}{2n-1-\sum_{j=1}^{n-1}p_{j-1}}
\frac{\widehat{\Sigma}_{0}^{
\mu_{I\setminus\bigcup_{s=1}^{n-1}\sigma_{p_s}}}}{n}\,,
\end{multline}
where the summation over $\sigma$ complies with the rules described in Proposition \ref{lem18}.
\end{theorem}

\subsection{Third order}
As an example, we calculate the results in the third order, which were first obtained in \cite{16}. For this, we must use formulas (\ref{rt9}), (\ref{rt11}), and (\ref{ra7}) adapted to this particular case:
\begin{equation}
\label{t1}
a_3(x,x)=
\sum\limits_{p_1=0}^{2}
\sum\limits_{p_2=0}^{4-p_1}
\sum\limits_{\sigma}
\frac{\widehat{\Sigma}_{4-p_1}^{\mu_{\sigma_{p_1}}}}{4-p_1}
\frac{\widehat{\Sigma}_{5-p_2}^{\mu_{\sigma_{p_2}}}}{5-p_2-p_1}
\frac{\widehat{\Sigma}_{0}^{
\mu_{I\setminus\bigcup_{s=1}^{2}\sigma_{p_s}}}}{3}.
\end{equation}
To simplify the calculations, we note that
\begin{equation}
\left.(x-y)^{\nu}F_{\nu\mu}(x)\right|_{y=x}=0,\,\,\,\,\,\,F_{\mu\mu}(x)=0,
\end{equation}
and the operator $\widehat{S}_l^{\mu\nu\tau\tau}$ is symmetric in the indices $\mu$ and $\nu$. Taking the last remarks into account, we see that a nonzero contribution is given by the terms with $p_1=p_2=0$ and $p_1=0,\,\,p_2=3$, and formula $(\ref{t1})$ hence becomes
\begin{equation}
\label{t2}
a_3(x,x)=
\frac{1}{60}\widehat{\Sigma}_{0}^{
\mu\mu\tau\tau\nu\nu}+\frac{1}{12}
\left(\widehat{\Sigma}_{2}^{\mu\mu\tau}
\widehat{\Sigma}_{0}^{\tau\nu\nu}+
\widehat{\Sigma}_{2}^{\mu\tau\tau}
\widehat{\Sigma}_{0}^{\mu\nu\nu}\right).
\end{equation}
The formulas (\ref{rt6}) and (\ref{rt9}) implies that
\begin{multline}
\label{t3}
\widehat{\Sigma}_{0}^{
\mu\mu\tau\tau\nu\nu}=
\frac{4}{3}\nabla_{\mu}F_{\tau\nu}(x)\nabla_{\mu}F_{\tau\nu}(x)+
\frac{2}{3}\nabla_{\mu}F_{\mu\nu}(x)\nabla_{\tau}F_{\tau\nu}(x)+\\+
F_{\tau\nu}(x)\nabla_{\mu}\nabla_{\mu}F_{\tau\nu}(x)+
\nabla_{\mu}\nabla_{\mu}F_{\tau\nu}(x)F_{\tau\nu}(x)-
2F_{\tau\mu}(x)F_{\mu\nu}(x)F_{\nu\tau}(x)
\end{multline}
and
\begin{equation}
\label{t4}
\widehat{\Sigma}_{2}^{\mu\mu\tau}
\widehat{\Sigma}_{0}^{\tau\nu\nu}+
\widehat{\Sigma}_{2}^{\mu\tau\tau}
\widehat{\Sigma}_{0}^{\mu\nu\nu}=
-\frac{1}{15}\nabla_{\mu}F_{\mu\nu}(x)\nabla_{\tau}F_{\tau\nu}(x).
\end{equation}
After expressions $(\ref{t3})$ and $(\ref{t4})$ are substituted in
$(\ref{t2})$, the final formula hence becomes
\begin{multline}
\label{t5}
a_3(x,x)=
\frac{1}{45}\nabla_{\mu}F_{\tau\nu}(x)\nabla_{\mu}F_{\tau\nu}(x)+
\frac{1}{180}\nabla_{\mu}F_{\mu\nu}(x)\nabla_{\tau}F_{\tau\nu}(x)+\\+
\frac{1}{60}F_{\tau\nu}(x)\nabla_{\mu}\nabla_{\mu}F_{\tau\nu}(x)+
\frac{1}{60}\nabla_{\mu}\nabla_{\mu}F_{\tau\nu}(x)F_{\tau\nu}(x)-
\frac{1}{30}F_{\tau\mu}(x)F_{\mu\nu}(x)F_{\nu\tau}(x);
\end{multline}
which gives the standard answer after taking the trace and integrating.

\section{Appendix: Inverse P-exponential} 
Let $\chi_{x_1\geqslant\ldots\geqslant x_k}$ be the indicator function of the set
$$\{(x_1,\ldots,x_k)\in\mathbb{R^k}:x_1\geqslant\ldots\geqslant x_k\}$$ for $k>1$ and unit
for $k=1$. Let the function $\aleph$ be given by the formula
\begin{equation}
\label{c15}
\aleph(x_1,\ldots,x_k)=
\sum\limits_{n=1}^{k}\sum\limits_{k_1+\ldots+k_n=k}(-1)^n\chi_{x_1\geqslant\ldots\geqslant x_k}\ldots\chi_{x_{k_1+\ldots+k_{n-1}+1}\geqslant\ldots\geqslant x_{k_n}}.
\end{equation}
The following lemmas can be proved by induction.\\
\textbf{Lemma 1:} Let $c_n(x_1,\ldots,x_n)=\chi_{x_1\geqslant\ldots\geqslant x_n}$ for $n\geqslant1$ and $c_0=1$, and for $n\geqslant1$, let
\begin{equation}
\label{c18} 
d_{n}(x_1,\ldots,x_n)=-c_n(x_1,\ldots,x_n)-\sum\limits_{k=1}^{n-1}d_{n-k}(x_1,\ldots,x_{n-k})c_k(x_{n-k+1},\ldots,x_n).
\end{equation}
Then $d_{n}(x_1,\ldots,x_n)=\aleph(x_1,\ldots,x_n)$ для $n\geqslant1$.\\
\textbf{Lemma 2:} We have the equality $\aleph(x_1,\ldots,x_n)=(-1)^n\chi_{x_1\leqslant\ldots\leqslant x_n}$.\\
\textbf{Proof:} For $k=1$, the formula is obvious. We suppose that it holds for $n=k-1$. Then for $n=k$,
\begin{equation*}
\aleph(x_1,\ldots,x_n)=-\left(\sum\limits_{k=1}^{n}(-1)^{n-k}\chi_{x_1\leqslant\ldots\leqslant x_{n-k}}\chi_{x_{n-k+1}\geqslant\ldots\geqslant x_n}\right).
\end{equation*}
But
\begin{equation*}
\chi_{y_1\leqslant\ldots\leqslant y_{n-1}}\chi_{y_n}-\chi_{y_1\leqslant\ldots\leqslant y_{n-2}}\chi_{y_{n-1}\geqslant y_n}=\chi_{y_1\leqslant
\ldots\leqslant y_{n}}-\chi_{y_1\leqslant\ldots\leqslant y_{n-2}}\chi_{y_{n-2}\geqslant\ldots\geqslant y_n},
\end{equation*}
and for $2\leqslant j\leqslant n-1$,
\begin{equation*}
\chi_{y_1\leqslant\ldots\leqslant y_{n-j}}\chi_{y_{n-j}\geqslant\ldots\geqslant y_n}-\chi_{y_1\leqslant\ldots\leqslant y_{n-j-1}}
\chi_{y_{n-j}\geqslant\ldots\geqslant y_n}=\chi_{y_1\leqslant\ldots\leqslant y_{n-j-1}}\chi_{y_{n-j-1}\geqslant\ldots\geqslant y_n},
\end{equation*}
which completes the proof. $\blacksquare$\\
To determine the inverse operator, we can regard (\ref{b18}) as a series in the background field. Indeed, after the transformation
$B_{\mu}(y)\rightarrow pB_{\mu}(y)$, where $p$ is a certain variable, the operator can be expanded in a series
\begin{equation}
\label{c7}
\Phi(x,y)=1+\sum\limits_{n=1}^{\infty}p^na_n,\,\,\,\,\,\,a_n=a_n(B,x,y).
\end{equation}
The ansatz for the inverse operator has a similar form:
\begin{equation}
\label{c8}
\Phi^{-1}(x,y)=1+\sum\limits_{n=1}^{\infty}p^nb_n,\,\,\,\,\,\,b_n=b_n(B,x,y). 
\end{equation}
The relation $\Phi^{-1}(x,y)\Phi(x,y)=1$ implies the system of recurrence relations
\begin{equation}
\label{c12}
b_{n}=-a_n-\sum\limits_{k=1}^{n-1}b_{n-k}a_k,\,\,n\geqslant1,
\end{equation}
whence we take Definition \ref{def1} and Lemmas 1 and 2 into account and obtain $\Phi^{-1}(x,y)=\Phi(y,x)$.

\section{Acknowledgments}
The author thanks S. E. Derkachev, T. A. Bolokhov, D. V. Vassilevich, and N. V. Kharuk for the discussions and the valuable comments.
This research was supported by a grant from the Russian Science Foundation (Project No. 14-11-00598).

\end{document}